\pdfoutput=1

\documentclass[11pt,twoside,a4paper,cmspaper,final,collab]{cms-tdr}
\def\svnVersion{227782}\def\cmsCernNo{2013-204}\def\cmsCernDate{\today}\def\cmsMessage{Published in Physics Letters B as \href{http://dx.doi.org/10.1016/j.physletb.2014.01.049}{\doi{10.1016/j.physletb.2014.01.049.}}}

\begin{document}\cmsNoteHeader{EXO-12-049}

\hyphenation{had-ron-i-za-tion}
\hyphenation{cal-or-i-me-ter}
\hyphenation{de-vices}

\RCS$Revision: 224443 $
\RCS$HeadURL: svn+ssh://svn.cern.ch/reps/tdr2/papers/EXO-12-049/trunk/EXO-12-049.tex $
\RCS$Id: EXO-12-049.tex 224443 2014-01-24 16:42:45Z alverson $
\newlength\cmsFigWidth
\ifthenelse{\boolean{cms@external}}{\setlength\cmsFigWidth{0.98\columnwidth}}{\setlength\cmsFigWidth{0.6\textwidth}}
\newlength\cmsFigWidthTwo
\ifthenelse{\boolean{cms@external}}{\setlength\cmsFigWidthTwo{0.98\columnwidth}}{\setlength\cmsFigWidthTwo{0.49\textwidth}}
\ifthenelse{\boolean{cms@external}}{\providecommand{\cmsLeft}{top}}{\providecommand{\cmsLeft}{left}}
\ifthenelse{\boolean{cms@external}}{\providecommand{\cmsRight}{bottom}}{\providecommand{\cmsRight}{right}}
\cmsNoteHeader{EXO-12-049} 
\title{Searches for light- and heavy-flavour three-jet resonances in
pp collisions at $\sqrt{s} = 8\TeV$}

\date{\today}

\abstract{
A search for three-jet hadronic resonance production in pp collisions at a
centre-of-mass
energy of 8\TeV has been conducted by the CMS Collaboration at the LHC with
a data sample corresponding to an integrated luminosity of 19.4\fbinv.
The search method is model independent, and
events are selected that have
high jet multiplicity and large values of jet transverse momenta.
The signal
models explored assume $R$-parity-violating supersymmetric gluino pair
production and have final states with either only light-flavour jets or both light-
and heavy-flavour jets. No significant deviation is found between the selected
events and the expected standard model multijet and \ttbar background.
For a gluino decaying into light-flavour jets, a lower limit of 650\GeV on the gluino mass is
set at a 95\% confidence level,
and for a gluino decaying into one heavy- and two light-flavour jets, gluino
masses between 200 and 835\GeV are, for the first time, likewise excluded.
}

\hypersetup{%
pdfauthor={CMS Collaboration},%
pdftitle={Searches for light- and heavy-flavour three-jet resonances in pp collisions at sqrt(s) = 8 TeV},%
pdfsubject={CMS},%
pdfkeywords={CMS, physics, software, computing}}

\maketitle 

\section{Introduction}
Hadronic multijet final states at hadron colliders offer a
unique window on many possible extensions of the standard model (SM),
although with the view partly obscured by large backgrounds due to SM processes.
Many of these extensions predict resonances, such as
heavy coloured fermions transforming as octets under
$SU(3)_\text{c}$
~\cite{tcgut,exoquarks,unichiral,altchiral} or supersymmetric gluinos
that undergo $R$-parity-violating (RPV) decays to
three quarks~\cite{sixjets,multijets,reuven}.
Recent studies from the Fermilab Tevatron Collider and 
the CERN Large Hadron Collider (LHC) employed the jet-ensemble technique. 
For this technique, jets are associated into unique 
combinations of three jets (triplets). Additional selection requirements 
are imposed to suppress the large backgrounds due to SM processes and 
to enhance sensitivity to strongly decaying resonances.
These analyses set lower mass limits based upon resonance fits for gluinos undergoing RPV decays.
The CDF collaboration at the Tevatron excluded gluino masses below 144\GeV~\cite{cdfmultijets}
using data from p$\bar{\textrm{p}}$ collisions at 1.96\TeV, while
the CMS collaboration at the LHC excluded masses
below 460\GeV~\cite{cmsmultijets, gluino2011} with data from pp collisions at 7\TeV.
An additional search at the LHC by the ATLAS collaboration, also based on data collected with
pp collisions at 7\TeV, has extended these limits to
666\GeV~\cite{gluinoATLAS}.

Presented here are the results of dedicated searches for pair-produced three-jet
resonances in multijet events from
\Pp\Pp\ collisions, with one search being inclusive with respect to parton flavours
 and the second requiring at least one jet from
the resonance decay to be identified as a bottom-quark jet (b jet).
This latter, heavy-flavour search is the first of its kind and probes additional RPV couplings. The
results are based on a data sample of pp collisions at
$\sqrt{s}=8$\TeV, corresponding to an integrated luminosity of
$19.4\pm0.5$\fbinv~\cite{lumi} collected with the CMS detector~\cite{cmsdet} at the
LHC in 2012. Events with at least six jets, each with high
transverse momentum (\pt) with respect to the beam direction, are selected and investigated for
evidence of three-jet resonances consistent with strongly coupled
supersymmetric particle decays. The event selection criteria are
optimised in the context of the gluino signal mentioned above~\cite{sixjets,multijets,reuven},
using a simplified model where the gluinos decay with a branching fraction of 100\% to quark jets.
However, the generic features of the selection criteria provide a model-independent 
basis that can be used when examining extensions
of the SM, since any exotic three-jet resonance with a narrow width, sufficient cross
section, and high-\pt jets would be expected to produce a significant bump on the
smoothly falling SM background of our search.
Additionally, low trigger thresholds and the application of b-jet
identification make it possible to use SM top quark-antiquark (\ttbar) events
to validate the analysis techniques.

 \section{The CMS experiment}
The central feature of the CMS apparatus~\cite{cmsdet}
is a superconducting solenoid of 6\unit{m} internal diameter, providing a
magnetic field of 3.8\unit{T}. Within the superconducting solenoid volume
are a silicon pixel and strip tracker, a lead tungstate electromagnetic
calorimeter (ECAL), and a hadron calorimeter (HCAL) that consists of
brass layers and scintillator sampling calorimeters.
Muons are measured in gas ionisation detectors embedded in the steel
return yoke outside the solenoid.
Extensive forward calorimetry complements the coverage provided by the barrel and
endcap detectors.
CMS uses a right-handed coordinate system, with the origin at the nominal interaction point,
the $x$ axis pointing to the centre of the LHC, the $y$ axis pointing up
(perpendicular to the LHC plane), and the $z$ axis along the anticlockwise-beam direction.
The polar angle $\theta$ is measured with respect to the positive $z$ axis, the
azimuthal angle $\phi$ is measured in the $x$-$y$ plane, and
the pseudorapidity $\eta$ is defined as $\eta=- \ln[\tan(\theta/2)]$.
Energy deposits from hadronic
jets are measured using the ECAL and HCAL.
The energy resolution for photons with \ET ${\approx}$ 60\GeV varies
between 1.1\% and 2.6\% over the solid angle of the ECAL barrel, and
from 2.2\% to 5\% in the endcaps. The HCAL, when combined with the ECAL,
measures jets with a resolution
$\Delta E/E \approx 100\% / \sqrt{E\,[\GeVns]} \oplus 5\%$~\cite{Chatrchyan:2013dga}.
The ECAL provides coverage in pseudorapidity $\abs{ \eta }< 1.479 $ in a barrel region
and $1.479 <\abs{ \eta } < 3.0$ in two endcap regions.
In the region $\abs{ \eta }< 1.74$, the HCAL cells have widths of 0.087 in
$\eta$ and 0.087 in $\phi$. In the $\eta$-$\phi$ plane,
and for $\abs{\eta}< 1.48$, the HCAL cells map on to $5 \times 5$ ECAL crystals arrays to
form calorimeter towers projecting radially outwards from close to the
nominal interaction point.
At larger values of $\abs{ \eta }$, the size of the towers increases, and the
matching ECAL arrays contain fewer crystals. Within each tower, the energy deposits in
ECAL and HCAL cells are summed to define the calorimeter tower energies,
subsequently used to provide the energies and directions of hadronic jets.

The CMS detector uses a two-tier trigger system to collect data.
Events satisfying the requirements at the first level
are passed to the high-level trigger (HLT), 
whose output is recorded and limited to a total rate of ${\sim}350$\unit{Hz}.
An HLT requirement based on at least six jets,
reconstructed with only calorimeter information, is used to select events.
With the jets ordered in descending \pt values, the  \pt threshold at the HLT for the fourth jet is
60\GeV and, for the sixth jet, 20\GeV.
For events passing all offline requirements described
in Section~\ref{sec:evtsel}, the total trigger efficiency is at least 99\%.

The CMS particle-flow algorithm~\cite{pflow} combines calorimeter information with
reconstructed tracks to identify
individual particles such as photons, leptons, and neutral
and charged hadrons. The photon energy is
obtained directly from calibrated measurements in the ECAL.
The energy of electrons is determined from a combination of the track momentum at the
primary interaction vertex~\cite{goodPV},
the corresponding ECAL cluster energy, and the
energy sum of all bremsstrahlung photons associated with the track in the offline reconstruction.
The muon
energy is obtained from the corresponding track momentum. The
energy for a charged hadron is determined from a combination of the track
momentum and the corresponding ECAL and HCAL energies, corrected for
zero-suppression effects and calibrated for the nonlinear response
of the calorimeters. Finally, the energy of a neutral hadron is
obtained from the corresponding calibrated ECAL and HCAL energies.
The particle-flow objects serve as input for jet reconstruction,
performed using the anti-\kt algorithm~\cite{ak5, fastjet1, fastjet2} with a distance parameter of 0.5.
The jet transverse momentum resolution is typically 15\% at $\pt=10\GeV$, 8\% at 100\GeV,
and 4\% at 1\TeV; when jet clustering is based only upon the calorimeter energies,
the corresponding resolutions are about 40\%, 12\%, and 5\%.

Jet energy scale corrections~\cite{jec} derived from data and Monte
Carlo (MC) simulation are applied to account for the nonlinear
and nonuniform response of the calorimeters.
In data, a small residual correction factor is included to correct for
differences in jet response between data and simulation.
The combined corrections are approximately 5--10\%,
and their corresponding uncertainties range from 1--5\%,
depending on the pseudorapidity and energy of the jet. Jet quality
criteria~\cite{jetid} are applied to remove
misidentified jets, which arise primarily from calorimeter noise.
In both data and simulated signal events, more than 99.8\%
of all selected jets satisfy these criteria.

\section{Signal event simulation}
Pair-produced gluinos are used to model the signal. Gluino production
and decay are simulated using the \PYTHIA~\cite{pythia} event generator~(v6.424),
with each gluino decaying to three quarks through
the $\lambda''_{\cPqu\cPqd\cPqd}$ quark RPV coupling~\cite{susycpling}, where $\cPqu$ and $\cPqd$
refer to any up- or
down-type quark, respectively. Two different scenarios
are considered for this coupling, resulting in both
an inclusive search similar to
previous analyses~\cite{cdfmultijets, cmsmultijets, gluino2011, gluinoATLAS}
and a new heavy-flavour search. For the first case,
the coupling of $\lambda''_{{112}}$,
where the three numerical subscripts of $\lambda$
refer to the quark generations of the corresponding u-d-d quarks,
is
set to a non-zero value, giving a branching fraction of 100\% for the gluino decay to
three light-flavour quarks.
The second case, represented
by $\lambda''_{\textrm{113}}$ or $\lambda''_{{223}}$, covers
gluino decays to one b quark and two light-flavour quarks.
The mass of the generated gluino signal ranges
from 200 to 500\GeV in 50\GeV steps,
with additional mass points at 750, 1000, 1250, and 1500\GeV.
For the generation of this signal, all superpartners
except the gluino are taken to be decoupled and heavy
(\ie beyond the reach of the LHC), the natural
width of the gluino resonance is taken to be much smaller than the
mass resolution of the detector of approximately 4--8\% in the mass range investigated,
and no intermediate particles are produced in
the gluino decay.
Simulation of the CMS detector response is performed
using the \GEANTfour~\cite{geant4} package.

\section{Event selection}
\label{sec:evtsel}
Events recorded with the six-jet trigger described above
are required
to contain at least one reconstructed primary vertex~\cite{goodPV}.
Since this analysis targets pair-produced three-jet resonances that
naturally yield high jet multiplicity, we
require events to contain at least six jets with $\abs{\eta} < 2.5$.
To ensure that the trigger is fully efficient, we impose minimal requirements
that the \pt thresholds of the fourth and sixth jets
are at least 80 and 60\GeV, respectively, though we impose higher thresholds for two of our three
selections, as described below.

\begin{figure}[htb]
 \begin{center}
   \includegraphics[width=\cmsFigWidth]{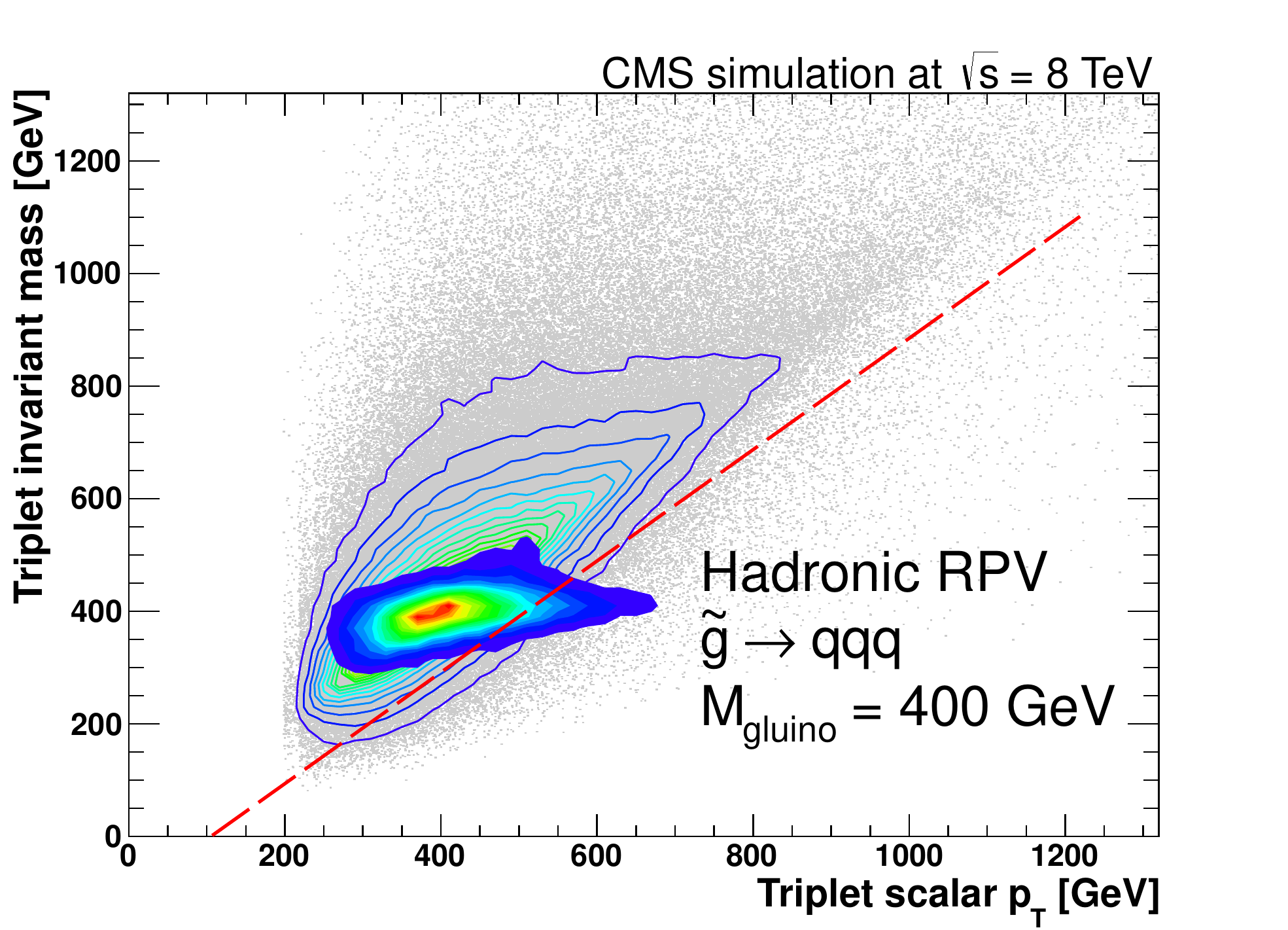}
   \caption{The triplet invariant mass versus the triplet scalar \pt
            for all combinations of the six jets from pair-produced gluinos
						of mass 400\GeV that decay to three light-flavour jets.
	    The solid coloured regions represent correctly reconstructed signal triplets,
	    while the contour lines and light grey scatter points represent incorrectly
	    combined triplets. The red dashed line is based on Eq.~(\ref{eq:diag_def}) with
	    $\Delta=110\GeV$,
	    and the triplets to the right of the line satisfy
	    this requirement, while those to the left do not.}
 \label{fig:MjjjvsSumpt}
 \end{center}
\end{figure}

We use the jet-ensemble technique~\cite{cdfmultijets, cmsmultijets}
in this analysis to combine the six highest-\pt jets
in each event into all possible unique triplets.
Each event that
satisfies all selection requirements will yield 20
combinations of jet triplets. For signal events, no more than
two of these triplets can be correct reconstructions of the
pair-produced gluinos,
with the remaining 18 triplets being incorrect combinations of jets.
Thus, background triplets arising from SM multijet events
are supplemented by ``incorrect'' jet-triplet combinations
from the signal events themselves. To obtain sensitivity to
the presence of a three-jet resonance, an additional requirement
is placed on each jet triplet to preferentially remove SM
background and incorrectly combined signal triplets. This selection
criterion exploits the constant invariant mass of correctly reconstructed signal triplets and
the observed linear correlation between the invariant mass and scalar sum of jet \pt
for background triplets and incorrectly combined signal triplets:
\begin{equation}
M_{jjj} < \left( \sum_{i=1}^{3}\pt^{i} \right)-\Delta \;,
\label{eq:diag_def}
\end{equation}
where $M_{jjj}$ is the triplet invariant mass,
the \pt sum is over the three jets in the triplet (triplet scalar \pt), and $\Delta$
is an empirically determined parameter.
Figure~\ref{fig:MjjjvsSumpt} shows a plot of the triplet invariant mass versus triplet scalar \pt for
simulated events with 400\GeV gluinos decaying to light-flavour jets.

The value of $\Delta$ is chosen so that the analysis is sensitive to as broad a range of gluino masses
as possible given the restrictions imposed by the trigger.
We find that the peak position of the $M_{jjj}$ distribution in data depends on the value of $\Delta$.
From a study of this peak position versus $\Delta$, we find $\Delta=110\GeV$ to
be the optimal choice, yielding the lowest value of the peak of $M_{jjj}$. This simple $\Delta$ requirement, rather than
model-specific invariant mass requirements, maintains the model-independent sensitivity of our
analysis to any three-jet resonance, not just that of our signal model.

Tightening the selection requirement on the \pt value of the sixth jet can reduce
background stemming from SM multijet production.
The optimisation of this requirement to maximise signal significance is performed as follows.

\begin{figure}[htbp]
\begin{center}
\includegraphics[width=\cmsFigWidth]{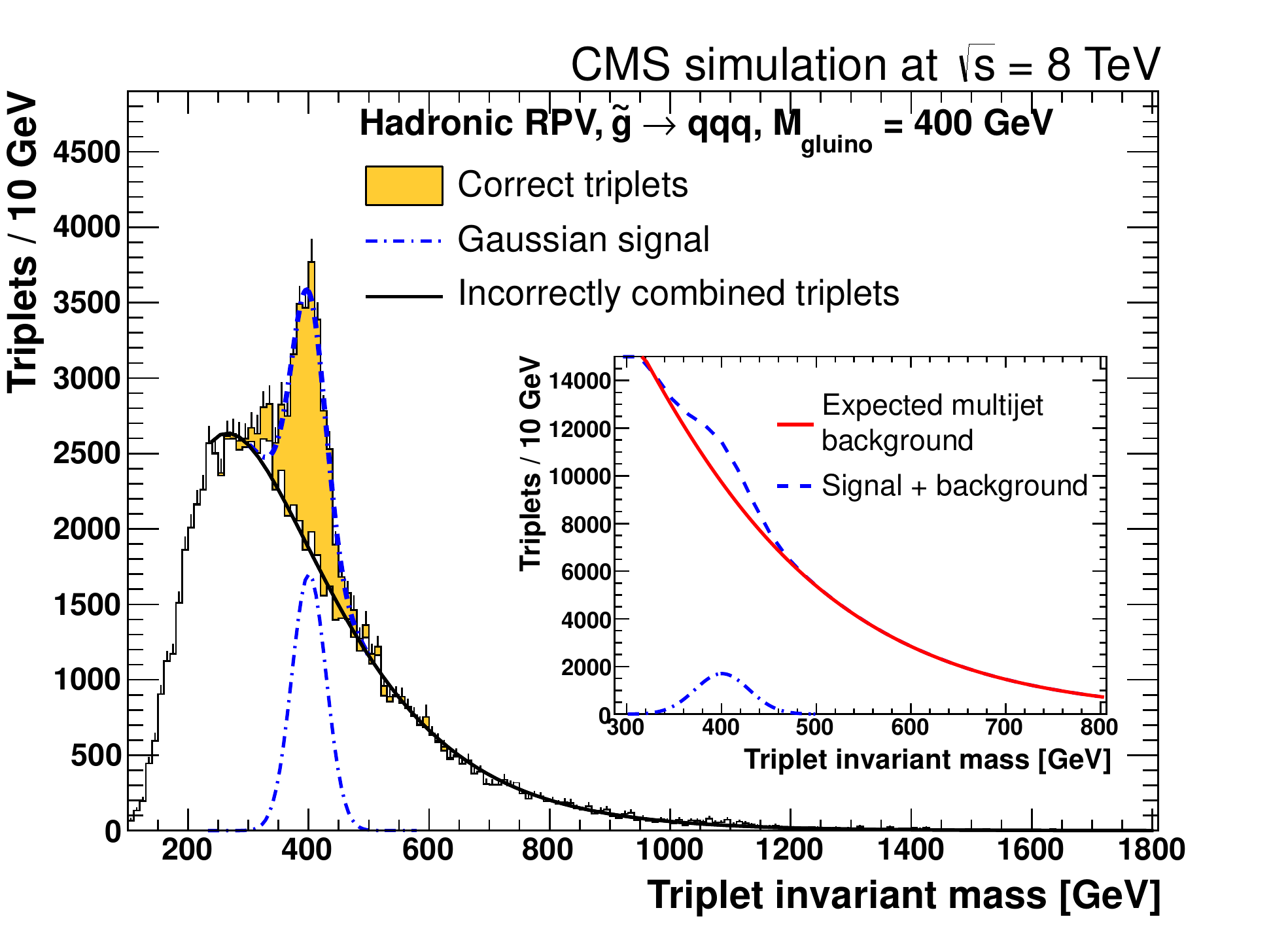}
\caption{The $M_{jjj}$~distribution for pair-produced 400\GeV
     gluinos with light-flavour RPV decay into three jets is shown in the main plot.
Triplets are selected that pass the $\Delta=110\GeV$ requirement from Eq.~(\ref{eq:diag_def}).
     The Gaussian signal peak of correctly reconstructed gluino triplets is represented by the gold
shaded area, with its Gaussian fit shown by the blue dot-dashed line below it. The distribution
     of incorrectly
     combined triplets, shown in black, is described by a similar functional form as that
		 used to estimate the background in data.
		 The inset shows the signal and background estimates used in the optimisation
		 procedure, with the expected background from SM multijet processes in red, and the
     signal-plus-background indicated by a blue dashed line.}
\label{fig:ParamsM400}
\end{center}
\end{figure}

As illustrated in Fig.~\ref{fig:ParamsM400} for a gluino mass of 400\GeV, the triplet invariant
mass distribution for signal events has the
shape of a Gaussian peak on top of a broad base of incorrect three-jet combinations.
We define the Gaussian peak to be the signal.
Following Ref.~\cite{dijetmass8tev}, we
use a four-parameter function (Eq.~(\ref{Eq:p4})) that is representative of the
estimated background in the data (see Section~\ref{sec:sigbkg})
and characterised by a steeply and monotonically decreasing shape:
\begin{equation}
\frac{\rd{}N}{\rd{}x}=P_0 \frac{\left(1 - \frac{x}{\sqrt{s}}\right)^{P_1}}{\left(\frac{x}{\sqrt{s}}\right)^{P_2 + P_3 \log{\frac{x}{\sqrt{s}}}}} \;,
\label{Eq:p4}
\end{equation}
where $N$ is the number of triplets and $x$ is the triplet invariant mass. The parametrised
signal and background estimates used in the optimisation procedure can be seen in the inset
of Fig.~\ref{fig:ParamsM400}.

Using these two components, signal triplets from the Gaussian peak and background triplets
from the background estimate, we define
the signal significance as the ratio of the number of signal triplets
to the square root of the
number of signal triplets plus the number of background triplets obtained from data.
The number of signal and
background triplets is calculated within a window around the mass peak
with a width corresponding to twice the expected gluino-mass resolution.
This procedure is repeated for different thresholds on the sixth-jet \pt in steps of 10\GeV, for a given gluino mass.
For the inclusive search, the focus is on masses that are higher than those previously excluded
by the jet-ensemble technique~\cite{gluino2011},
so the mass range of the search starts around 400\GeV.
We find that a requirement of $\pt\geq110\GeV$ on the sixth jet maximises the signal significance in this mass range.

The use of b-jet identification enables us to perform a heavy-flavour search in addition to our
inclusive search for three-jet resonances.  The
combined secondary vertex (CSV) algorithm~\cite{btagger} uses variables from
reconstructed secondary vertices along with track-based lifetime
information to identify b jets.
The tagging efficiency for b jets changes
with the \pt of the jet, ranging from 70\% for jets with 100~$\leq$ \pt $\leq$ 200\GeV
to 55\% for jets with $\pt \geq 500\GeV$.
We study different b-tagging requirements for
signal events with simulated gluinos that have heavy-flavour decays
and use the same definition of the signal significance as for the sixth-jet \pt
optimisation to determine the best choice.
The CSV medium operating point, with
a mistagging rate of about 1\% for light-flavour jets, is
found to be the optimal choice for detecting a potential signal
in this analysis.
The requirement that
each event contain at least one b-tagged jet (b tag) increases the signal
significance, and the additional requirement that all selected
triplets have a b~tag removes a large portion of the incorrectly
combined signal triplets.

For the heavy-flavour analysis, we distinguish between a low-mass region
covering gluino masses between 200 and 600\GeV and a high-mass region
covering larger gluino masses. For the low-mass region,
we maximise signal acceptance by using
jet-\pt requirements of $\geq$80\GeV for the fourth jet and $\geq$60\GeV for the sixth jet.
For the high-mass region, the sixth jet is required to have \pt$\geq$110\GeV.
For both the low- and high-mass regions, the value $\Delta=110\GeV$ is used.
All-hadronic \ttbar event production is a significant background in the low-mass region.
We use \ttbar events that produce triplets with masses in this region to help validate our analysis technique, as described below.

\begin{figure}[htbp]
 \begin{center}
   \includegraphics[width=\cmsFigWidth]{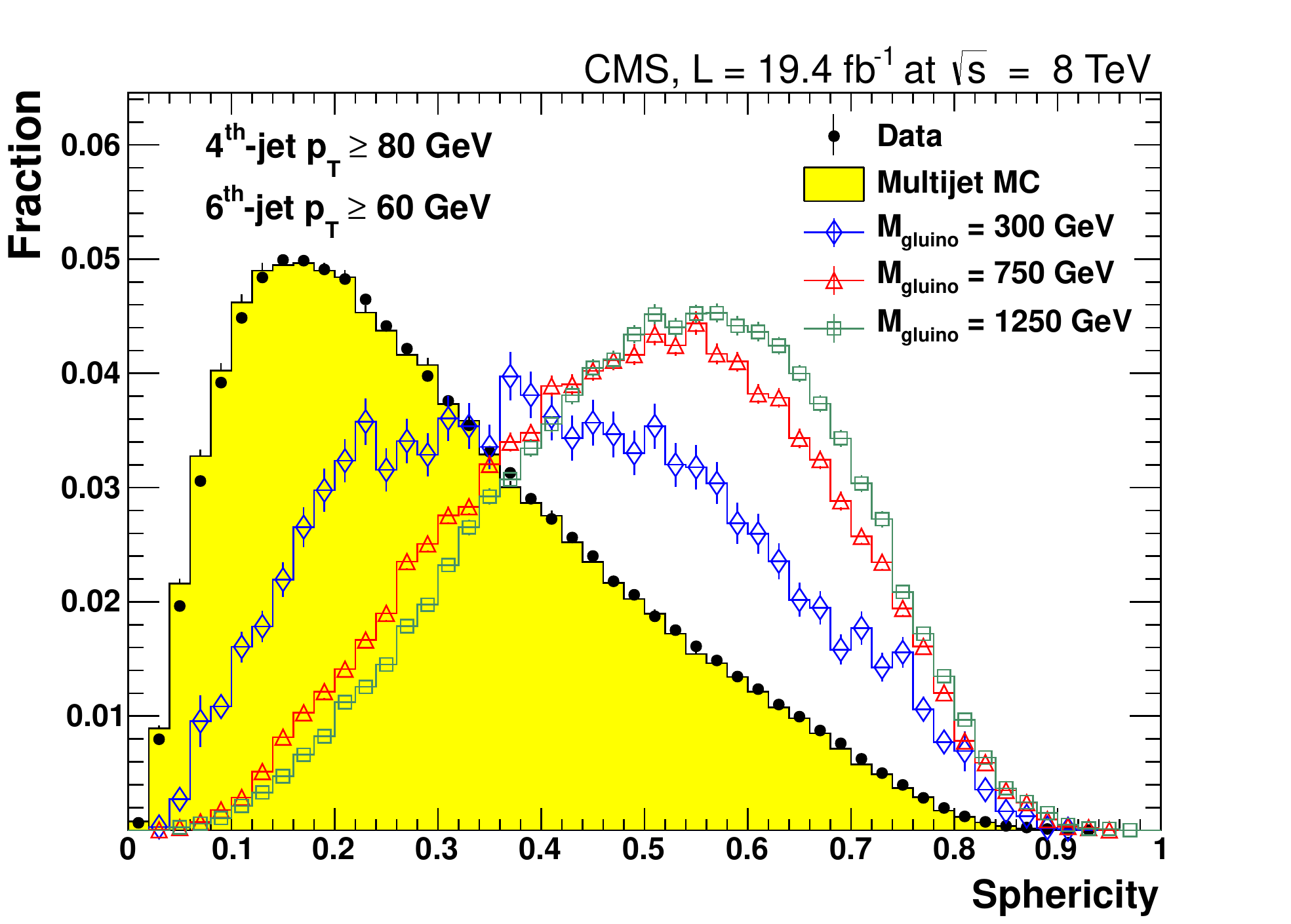}
   \caption{The sphericity variable for events from data, simulated background from SM multijet processes (shaded area),
            and simulated gluino signal masses of 300 (open diamonds), 750 (open triangles), and 1250\GeV (open squares),
 where the gluinos decay
						to light-flavour jets.
						Event-level selection requirements for the inclusive, low-mass search
						are applied, except for the triplet-level diagonal selection (Eq.~(\ref{eq:diag_def})).
						All distributions are normalised to unit area. The simulated SM multijet events
						are generated
						by \MADGRAPH~\cite{madgraph}
						with showering performed by \PYTHIA.}
   \label{fig:sphericity}
 \end{center}
\end{figure}

High-mass signal events, for both the light- and heavy-flavour signal models, have a
more spherical shape than background events, which typically
contain back-to-back jets and thus have a more linear shape.
To significantly reduce the background in the
high-mass searches, we use a sphericity
variable, $S = \frac{3}{2} (\lambda_{2}+\lambda_{3})$, where the
$\lambda_{i}$ variables are eigenvalues of the following tensor~\cite{pythia}:
\begin{equation}
S^{\alpha \beta} = \frac{\sum\limits_{i} p^{\alpha}_{i}  p^{\beta}_{i}}{\sum\limits_{i} \abs{\boldmath{p}_{i}}^2} \;,
\label{eqn:tensor}
\end{equation}
where $\alpha$ and $\beta$ label separate jets, and the sphericity $S$ is calculated
using all jets in each event.
A comparison of the sphericity variable for data, simulated SM multijet
events, and three different simulated gluino masses can be seen in
Fig.~\ref{fig:sphericity}. For the inclusive search and the
high-mass, heavy-flavour search, selected events are required to
have $S\geq0.4$, which is based on the optimisation of the number of expected signal
events divided by the square root of the number of signal-plus-background events.
No sphericity requirement is used for the low-mass, heavy-flavour
selection because low-mass signal events do not have a significant
shape difference from background events.

To conclude, we define three different search regions for this analysis with specific
selection criteria applied as previously discussed and summarised in Table~\ref{tab:cuts}.

\begin{table*}[htbp]
\begin{center}
\topcaption{Selection requirements for the three search regions in the analysis.}
\begin{tabular}{cccc}
\hline
 Selection & Inclusive & \multicolumn{ 2}{c}{Heavy-flavour search} \\
 criteria & search & low mass & high mass \\ \hline
Mass range & 400--1500\GeV & 200--600\GeV & 600--1500\GeV \\ \hline
$\Delta$ & 110\GeV & 110\GeV & 110\GeV \\
Min. fourth-jet \pt & 110\GeV & 80\GeV & 110\GeV \\
Min. sixth-jet \pt & 110\GeV & 60\GeV & 110\GeV \\
Min. sphericity & 0.4 & --- & 0.4 \\ \hline
\end{tabular}
\label{tab:cuts}
\end{center}
\end{table*}

\section{Background estimation and signal extraction}
\label{sec:sigbkg}
The dominant background for this search comes from SM multijet events, which
arise from perturbative QCD processes of order $\mathcal{O}(\alpha_s^3)$ and higher.
The invariant mass shape of incorrectly combined signal triplets
is found to be similar to that of the background from SM multijet processes, such that
the combined distribution is consistent with that of
SM multijets alone. Moreover, because the normalisation of the background component ($P_{0}$ in Eq.~(\ref{Eq:p4}))
is unconstrained, any incorrectly combined signal triplets, if present, would
be absorbed into the background estimate.
The triplet invariant mass distribution for the background
decreases smoothly with increasing mass, and we model this
background using a four-parameter
function (Eq.~(\ref{Eq:p4})) fit directly to the data,
except in the case of the low-mass, heavy-flavour search.

For the low-mass, heavy-flavour search, there is an
additional background contribution from all-hadronic \ttbar events.
These events are modelled using the \MADGRAPH~\cite{madgraph}
generator,
and the expected number of \ttbar events is determined from the
next-to-next-to-leading-order (NNLO) cross section of
$245.8 \substack{+8.7\\ -10.5}$\unit{pb}~\cite{ttbarxsec}. The
shape of the contribution from SM multijet processes is modelled with a statistically
independent data sample, constructed by imposing a veto on b-tagged jets
while retaining all other selection requirements. This sample is referred
to as the b-jet control region, and
the combination of simulated \ttbar events and the background from SM multijet processes,
modelled by this control region, gives
the total SM background estimate for the low-mass, heavy-flavour analysis.

\begin{figure}[htb]
 \begin{center}
   \includegraphics[width=\cmsFigWidth]{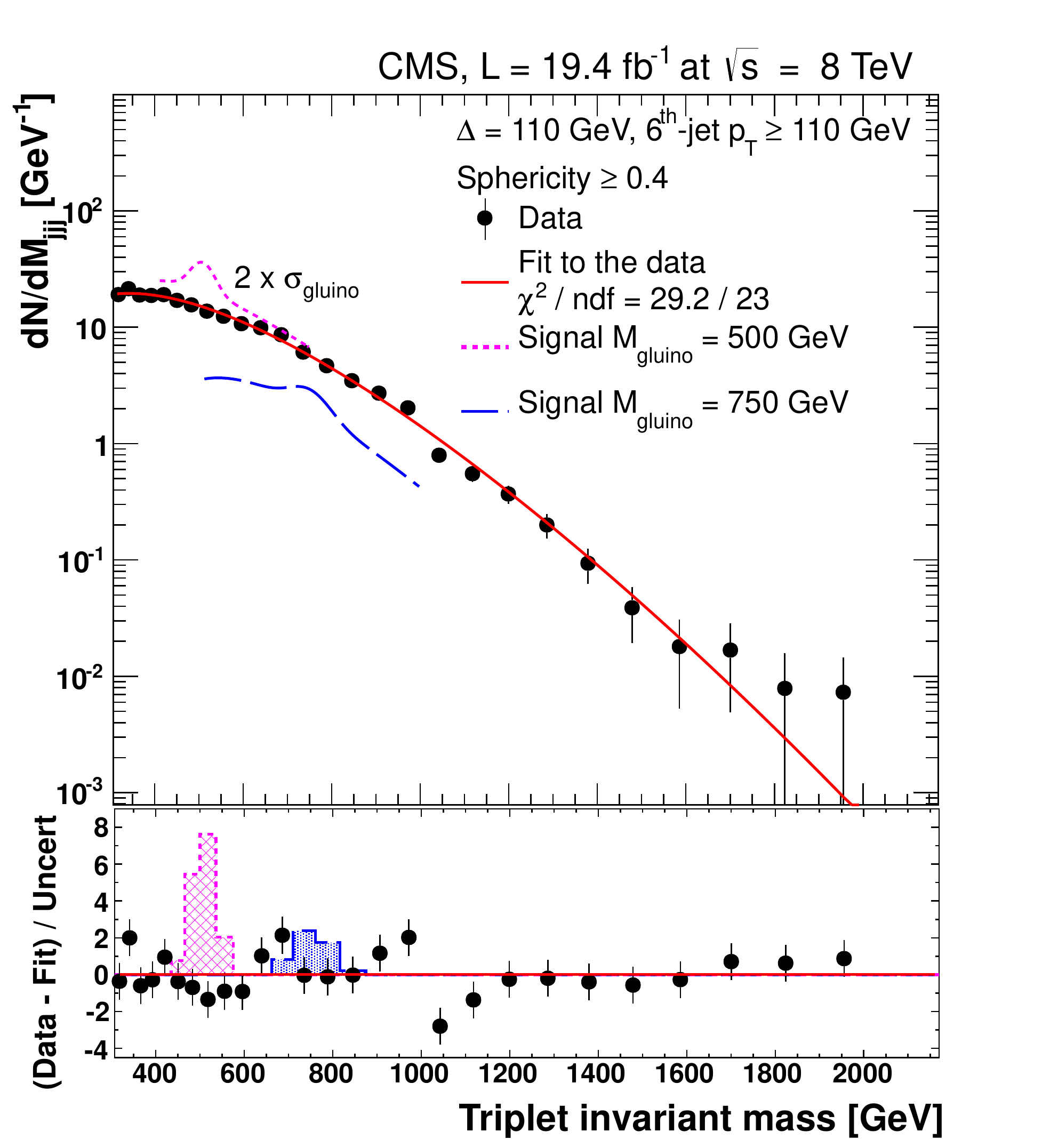}
   \caption{Comparison of the three-jet invariant mass distribution in data with the background estimate for the inclusive
	    analysis (red solid curve) obtained from
	    a maximum likelihood fit to the data. The error bars on the black data points 
display the statistical uncertainties. The bin widths increase with mass to match the expected resolution.
	    The bottom plot shows, for each bin, the difference of the data and fit values divided by the statistical
	    uncertainty in the data. No statistically significant deviations from the data are
	    observed.
The light magenta dotted line and hatched area show the distribution and pulls for a simulated 500\GeV
gluino that decays into light-flavour jets.
Similarly, the expectation for a 750\GeV gluino is shown by a dark blue dashed line and shaded area.
}
   \label{fig:DataBkgInc}
 \end{center}
\end{figure}

\begin{figure*}[htbp]
 \begin{center}
  \includegraphics[width=\cmsFigWidthTwo]{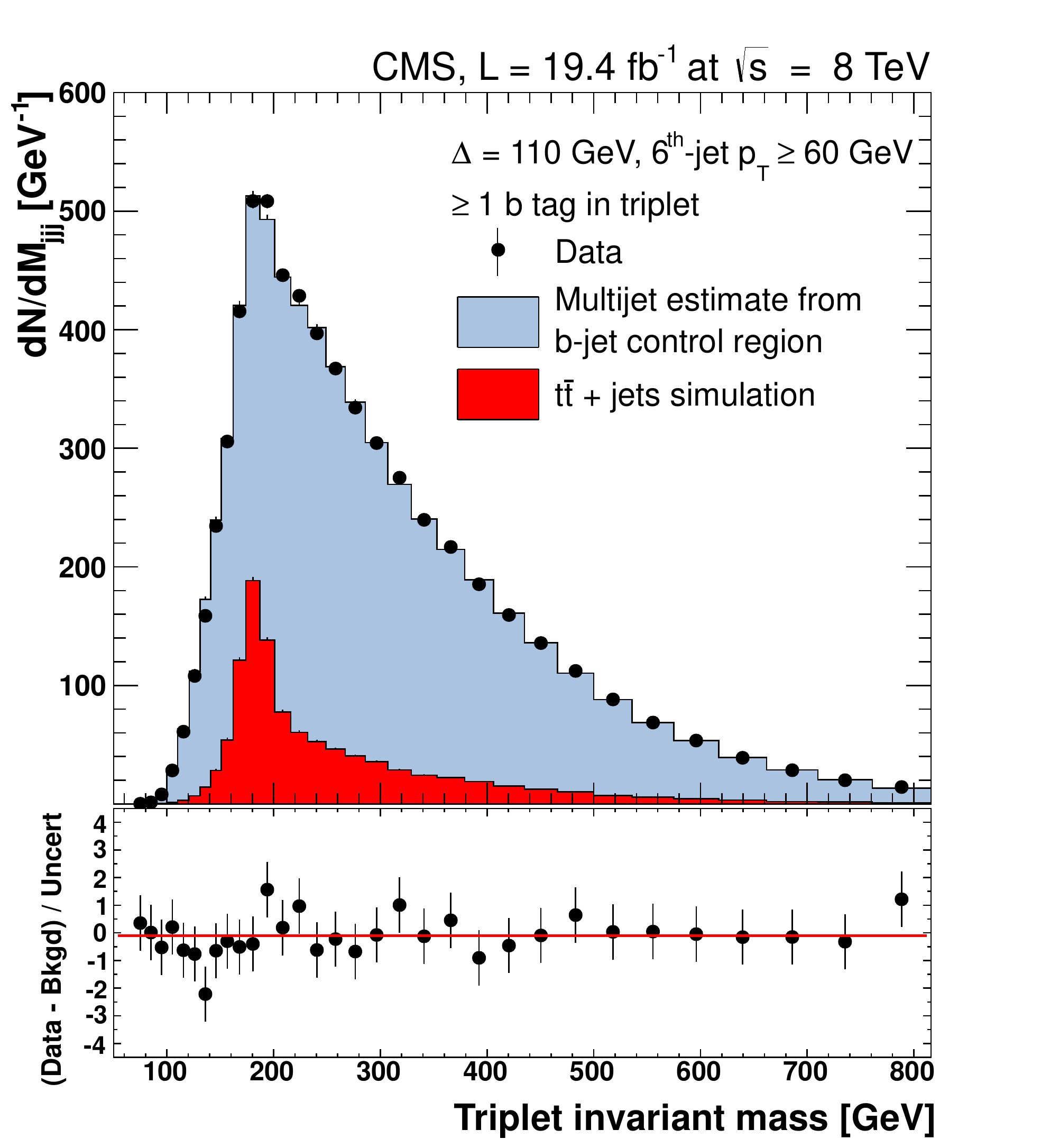}
   \includegraphics[width=\cmsFigWidthTwo]{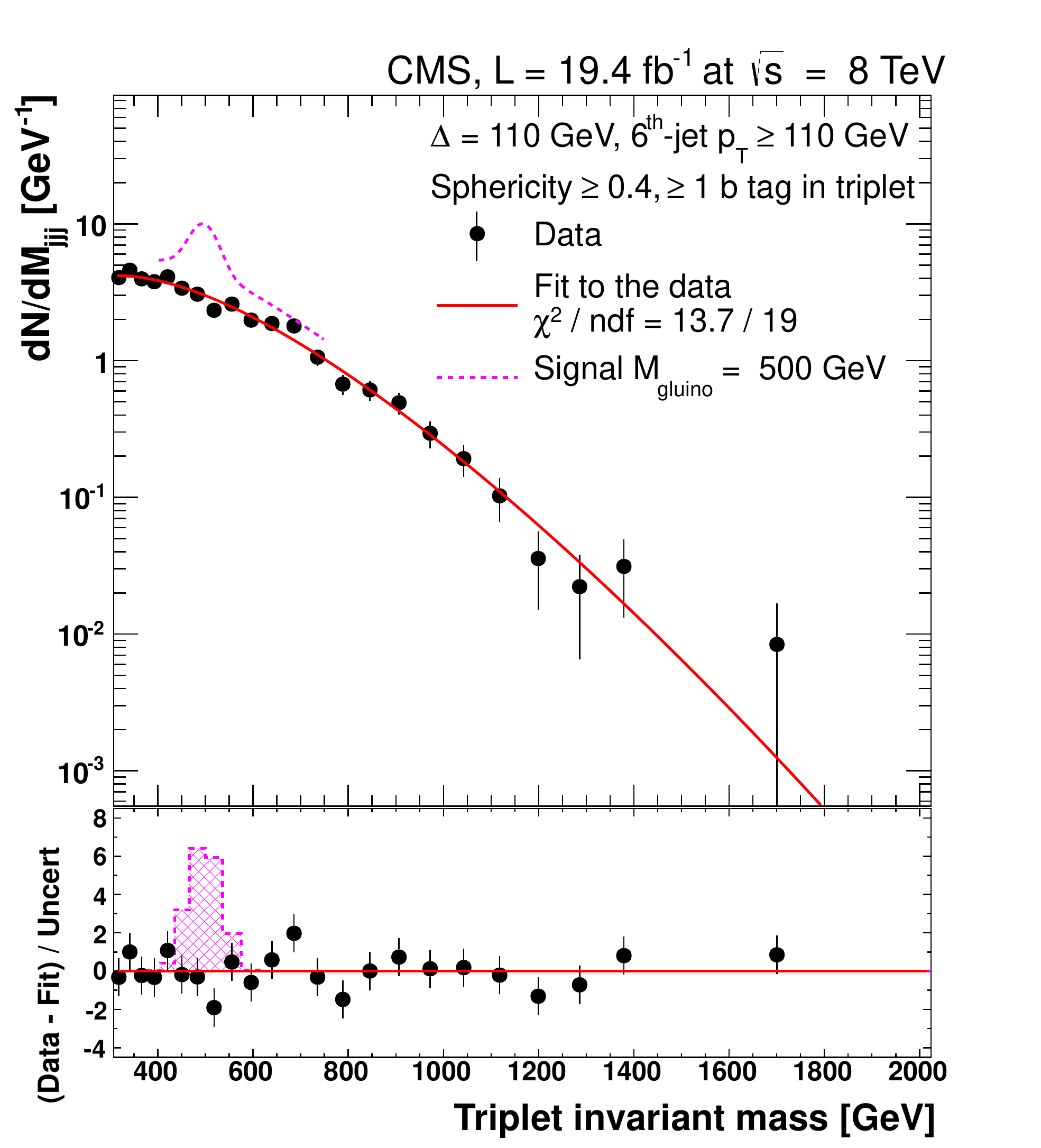}
   \caption{Comparison of the three-jet invariant mass distribution in data with the background estimate for the heavy-flavour
            analysis. The left plot shows the results from the low-mass selection.
            The background contribution from
	    the b-jet control region is shown as the light blue shaded area, while that from simulated \ttbar events is
	    shown as the dark red shaded area. The right plot shows the high-mass sample with resolution-based binning.
The error bars on the black data points
display the statistical uncertainties.
	    The bottom plots show the difference of the data and the background estimate divided by the statistical uncertainty in the data in each bin.
	    The light magenta dashed line and hatched area show
	    the distribution and pulls for a simulated 500\GeV gluino that decays
			into heavy-flavour jets.}
   \label{fig:DataBkg1b}
 \end{center}
\end{figure*}

A comparison of the background estimate to the data is
performed, in which the data are fit using a binned maximum likelihood method with
either the four-parameter function of Eq.~(\ref{Eq:p4}) for the inclusive analysis and
the high-mass, heavy-flavour analysis, or the background
shape described above for the low-mass, heavy-flavour analysis.
Figure~\ref{fig:DataBkgInc} shows a comparison between
the three-jet invariant mass distribution in data
and the background estimate for the inclusive analysis.
Figure~\ref{fig:DataBkg1b} shows the comparisons between data and
background estimates for the low- and high-mass heavy-flavour analyses.
In all three cases, no statistically significant deviations from the data are observed.

As a validation of the analysis technique, we consider the \ttbar triplets
as a signal with the background solely composed of triplets from SM multijet processes, whose shape
is modelled by the b-jet control region, with the small amount of
simulated \ttbar events without b tags subtracted.
The \ttbar cross section is extracted based on the contribution of its signal triplets
and is compared
with the theoretical prediction for the cross section of 245.8\unit{pb}.
The measurement yields a result of $205 \pm 28$\unit{pb} (combined
statistical and systematic uncertainties), which is within less than two
standard deviations from the theoretical value,
thereby showing our technique can successfully reconstruct hadronically decaying \ttbar events.

\begin{figure*}[htbp]
 \begin{center}
   \includegraphics[width=\cmsFigWidthTwo]{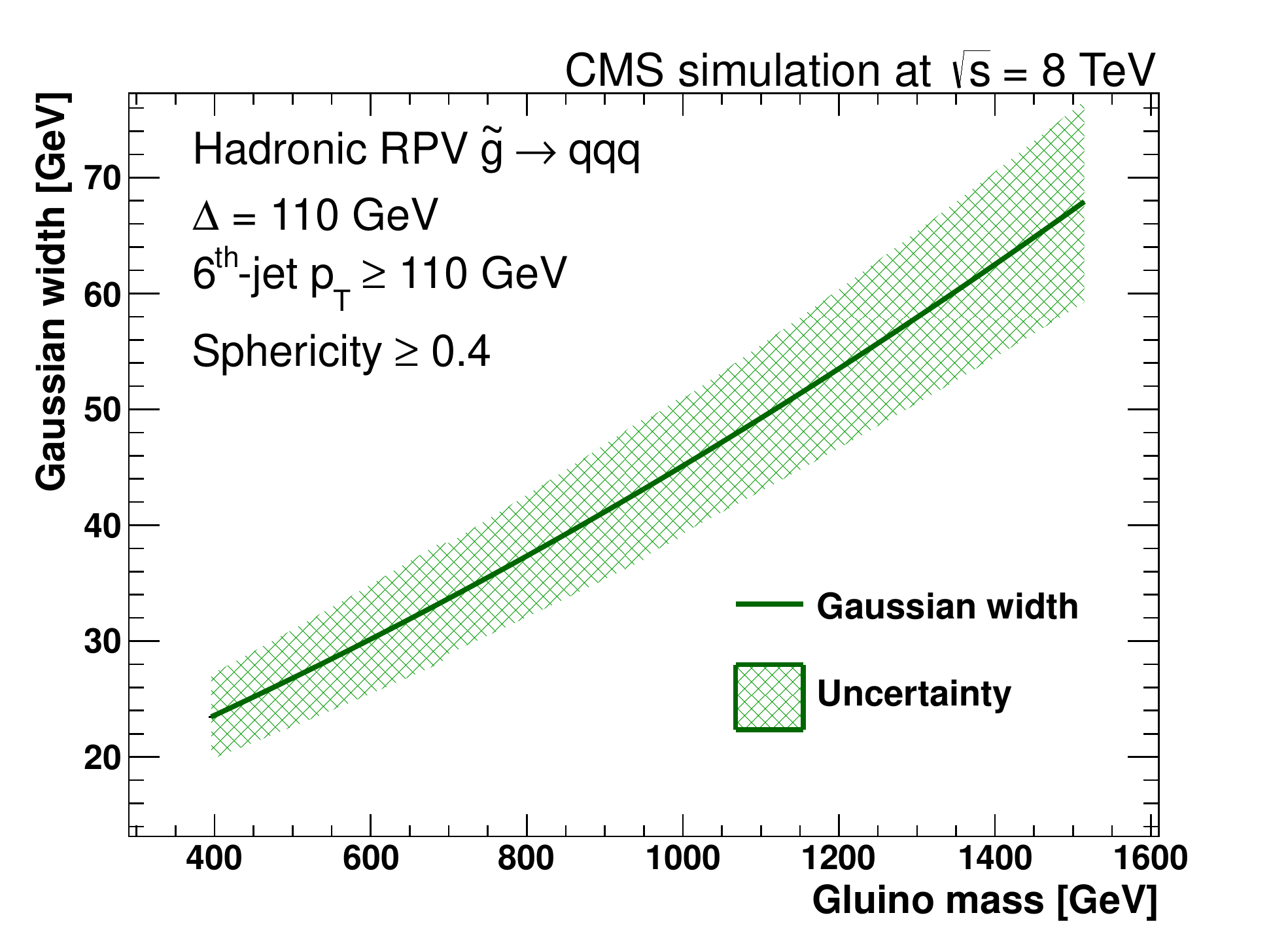}
   \includegraphics[width=\cmsFigWidthTwo]{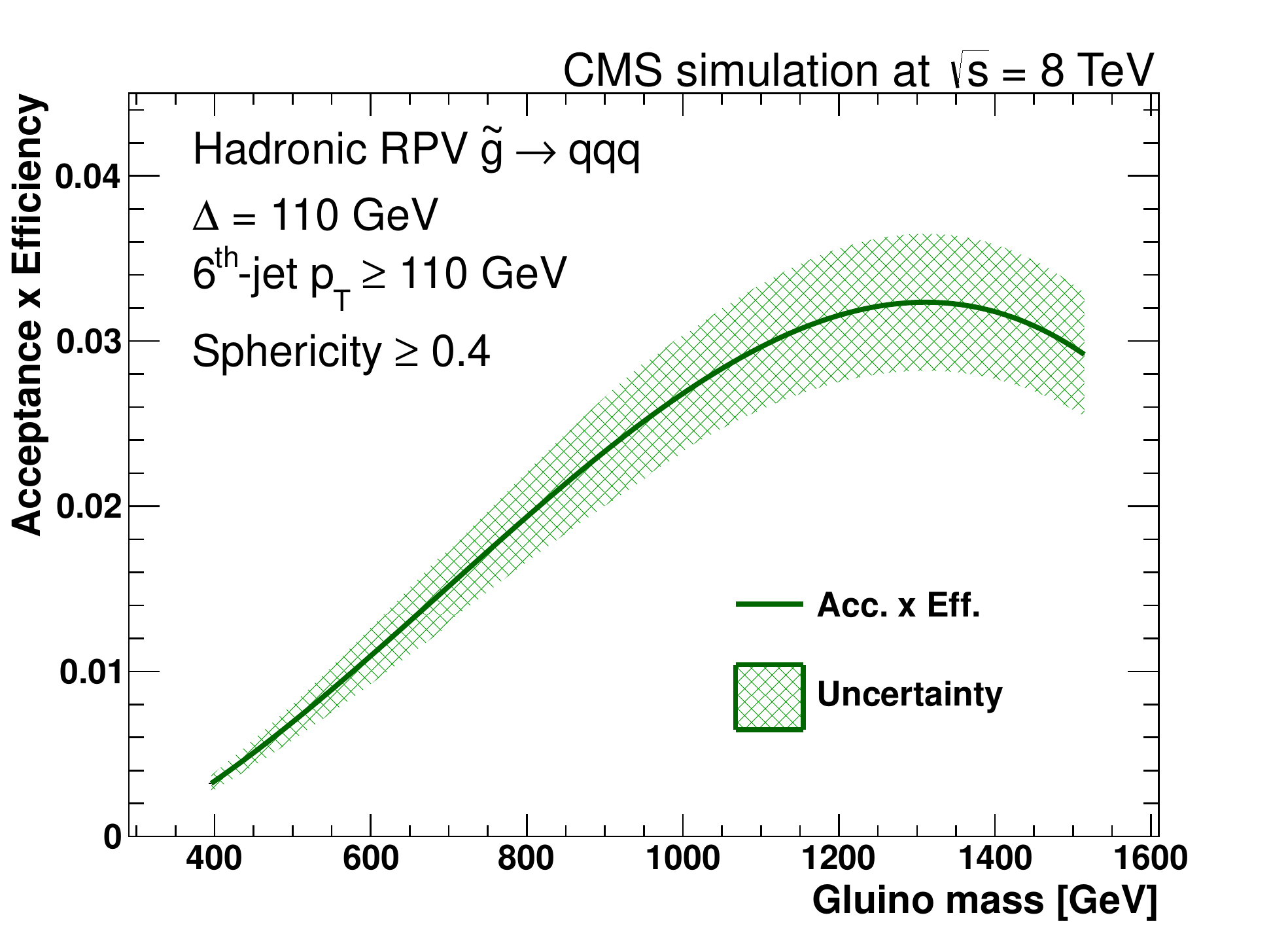}
   \caption{The signal parametrisation shown as a function of gluino mass for the inclusive search.
            The Gaussian signal width (left) and the $A \times \epsilon$
            parametrisation (right) versus signal mass are
	    both shown. The hatched bands represent the combined statistical and systematic uncertainties.}
   \label{fig:sigwidths}
 \end{center}
\end{figure*}

To obtain an estimate of the number of signal triplets expected after all selection
criteria are applied, the sum of a Gaussian function that represents
the signal and a four-parameter function (Eq.~(\ref{Eq:p4}))
that models the incorrectly combined signal triplets is fit to the simulated $M_{jjj}$
distribution for each gluino mass. The Gaussian
component of the fit provides the estimate for the expected number of signal triplets.
The factors in this overall triplet signal efficiency are the event acceptance, governed by the
kinematic and b-tagging selections, and the triplet rate, which represents
the number of selected triplets per selected event. This triplet rate
is the product of the average number of triplets per event
times the proportion of triplets contained in the Gaussian signal peak compared with
the full distribution. Width and acceptance-times-efficiency ($A \times \epsilon$)
are both parametrised as functions of gluino mass, as shown in Fig.~\ref{fig:sigwidths}.
The width of the Gaussian function modelling the signal varies according to the detector
resolution, ranging from 17 to 70\GeV for gluino masses from 200 to 1500\GeV.
The $A \times \epsilon$ ranges
from about 0.003 to 0.033 for the inclusive search for gluino masses from 400--1500\GeV, and, for the heavy-flavour search,
from 0.005 to 0.04 for masses from 200--600\GeV, and from 0.008 to 0.015 for
masses from 600--1500\GeV.
For high-mass gluinos, the $A \times \epsilon$ flattens slightly because of
the decreased efficiency to reconstruct triplets in the Gaussian signal peak.

\section{Systematic uncertainties}
Systematic uncertainties in the signal acceptance are assigned in the following manner. For uncertainties
related to the jet energy scale (JES)~\cite{jec}, the jet energy corrections are
varied within their uncertainties for each signal mass, and then the entire selection
procedure is repeated to determine the parametrised values of the
$A \times \epsilon$. The largest difference from the nominal values is taken as a
systematic uncertainty.
To evaluate the systematic uncertainty associated with the level of simulated ISR and FSR for signal events,
\ie the spontaneous emission of gluons from incoming or outgoing participants of the hard interaction,
dedicated signal samples are generated where the relative amounts of ISR and FSR are coherently
increased or decreased with respect to the nominal setting of the \PYTHIA event generator~\cite{isrfsruncert}.
The parameter controlling the amount of ISR (PARP(67)) is varied around its central value of 2.5 by ${\pm}0.5$ and
that for the FSR (PARP(71)) is varied from 2.5 to 8, with a nominal value of 4.0.
For each sample, the rederived $A \times \epsilon$ is compared to the nominal value, and
the difference is taken as the systematic uncertainty.
Analogously, an uncertainty is assigned to account for the effects of multiple pp
collisions in an event (pileup) by reweighting
all MC signal samples such that
the distribution of the number of interactions per bunch crossing is shifted,
high and low, by one standard deviation compared with that found in
data~\cite{pileupuncert}.
For the analyses using b tagging, an uncertainty is assigned based on the scale factor that
comprises the differences in b-tagging efficiencies in data compared with
simulation~\cite{btagger}.
The same procedure as outlined above is repeated, where the b-tagging scale factors are
varied within their uncertainties, and the effect on $A \times \epsilon$ is evaluated.
Uncertainties in the fit parameters of the Gaussian signal are used as an additional systematic
uncertainty for each mass point.
Finally, an overall systematic uncertainty of 2.6\% is assigned to the
integrated luminosity measurement~\cite{lumi}. The ranges in the values of these
uncertainties are summarised in Table~\ref{tab:sigsysunc}.
Systematic uncertainties related to the signal and background shapes are discussed in Section~\ref{sec:results}.
\begin{table}[htbp]
\begin{center}
\topcaption{Systematic uncertainties on the signal $A \times \epsilon$ included in limit setting.}
\begin{tabular}{cr}
\hline
Source of systematic uncertainty & Value  \\ \hline
JES & 3--16\%  \\
ISR/FSR & 5--11\%  \\
Pileup & 1--5\%   \\
b~tagging & 1--7\%  \\
Signal fit & 4--12\% \\
Luminosity &2.6\%  \\ \hline
\end{tabular}
\label{tab:sigsysunc}
\end{center}
\end{table}

\section{Results and limits}
\label{sec:results}

The three-jet invariant mass distributions are examined for a Gaussian signal peak on top of the smoothly falling background distribution.
As has been described, this analysis uses different selection criteria to search
for resonances coupling to light-flavour and to heavy-flavour quarks, with the
latter search done separately in low-mass and high-mass regions.
In the analysis of each of the three selections, the background normalisation parameter is unconstrained and is therefore determined by the SM multijet component of the combined fit.
For the function describing the background, the initial values of its 
parameters are taken from the background-only hypothesis fit
to the data, while they are allowed to float in the 
background-plus-signal hypothesis fits for the limit calculation.
The signal is modelled with Gaussians defined by the width and
$A \times \epsilon$ curves shown in Fig.~\ref{fig:sigwidths}. The
uncertainties in the expected number of signal triplets are included as log-normal constraints,
where the uncertainty for the width of the Gaussian includes a 10\% systematic uncertainty to account for
jet resolution effects~\cite{jec}.
For the \ttbar background estimate, uncertainties in both the shape and normalisation
are included. In addition to those already discussed in the previous section,
uncertainties due to ambiguities in the parton shower matching procedure between
the \MADGRAPH  and \PYTHIA event generators, as well as those due to the dependence
on the renormalisation and factorisation scale, are taken into account.

\begin{figure}[hbt]
\centering
\includegraphics[width=\cmsFigWidthTwo]{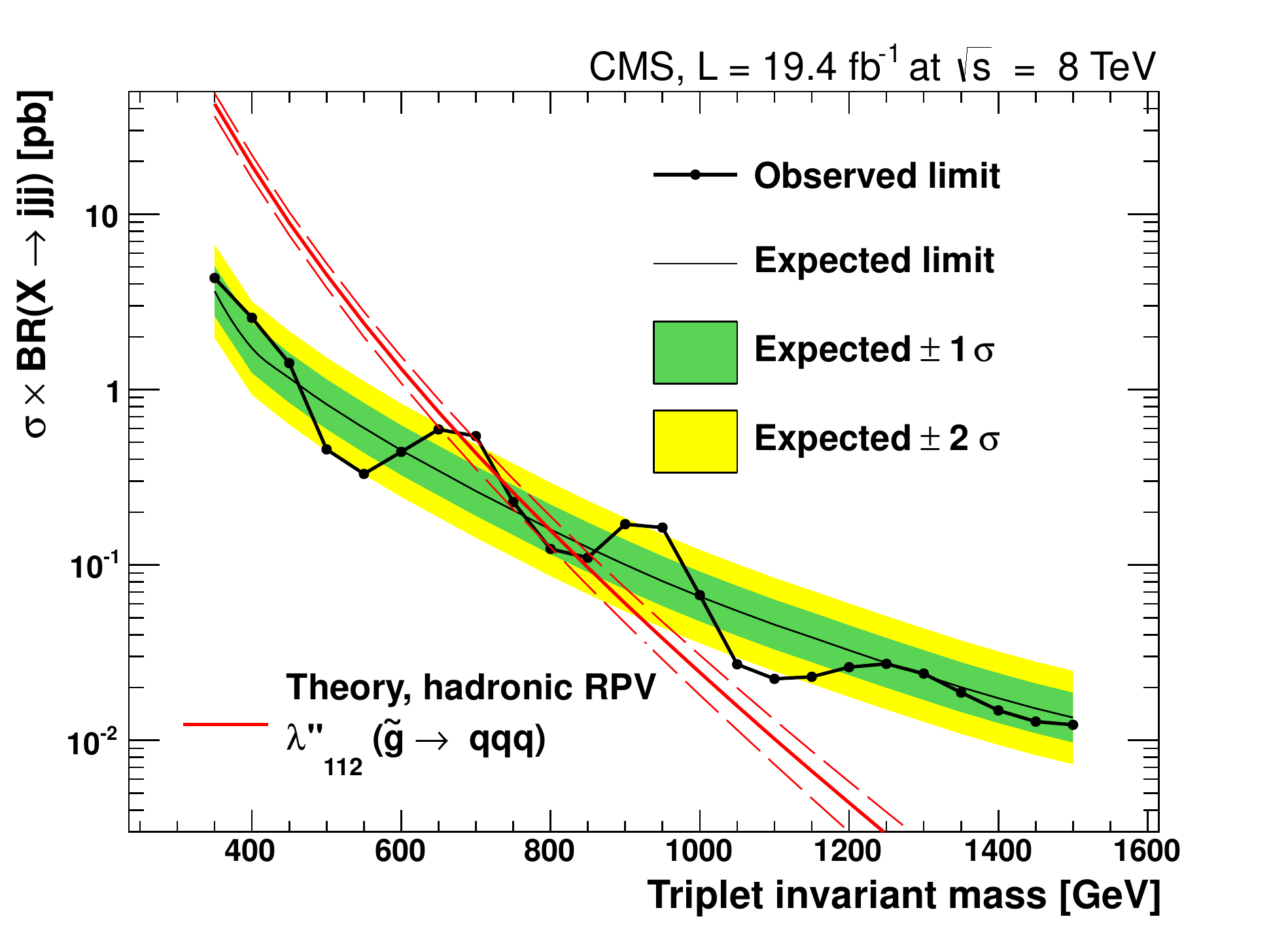}
\includegraphics[width=\cmsFigWidthTwo]{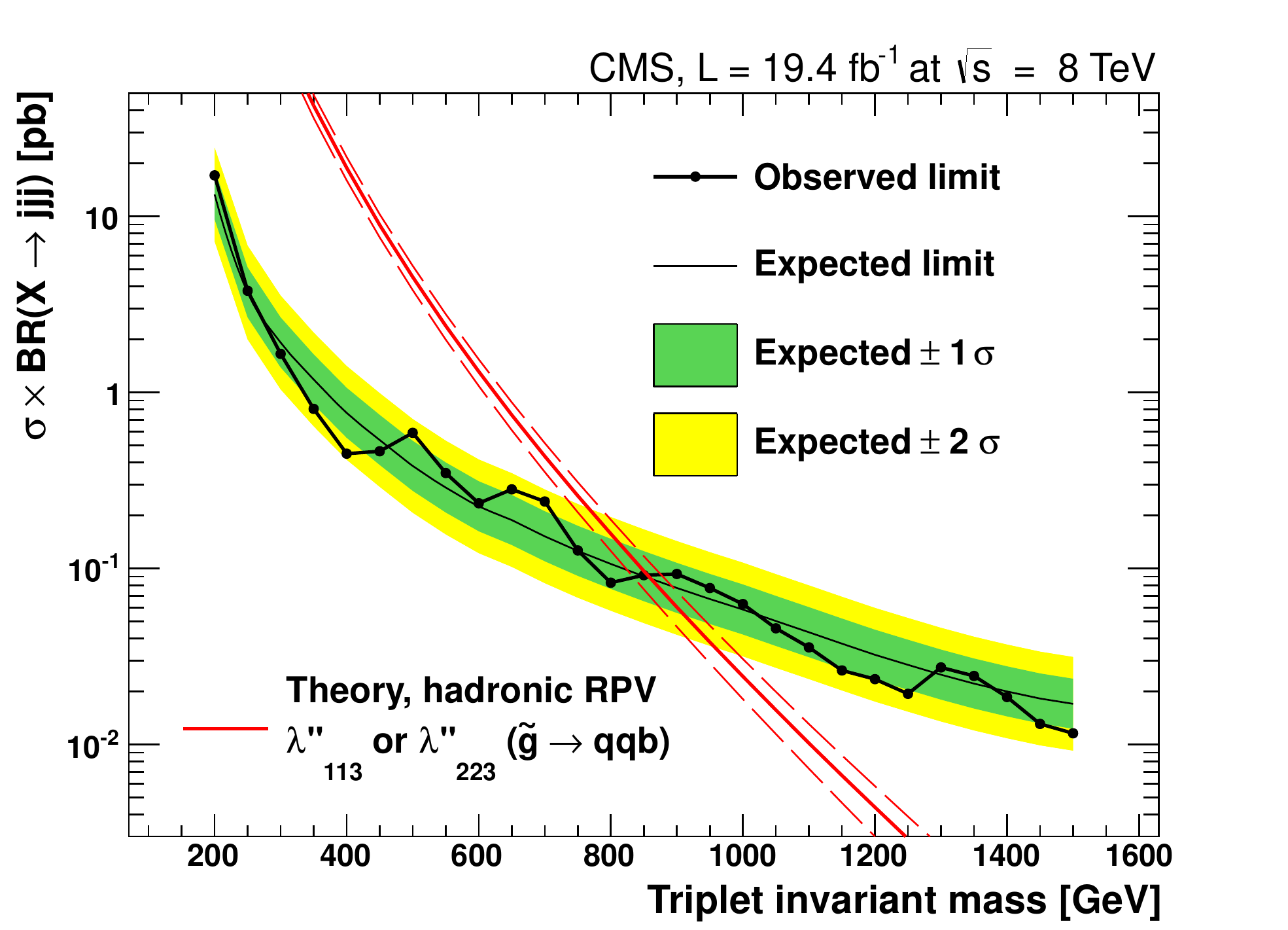}
\caption{Observed and expected 95\% CL cross section limits as a function of mass
         for the inclusive (\cmsLeft) and heavy-flavour searches (\cmsRight).
	  The limits for the heavy-flavour
	 search cover two mass ranges, one for low-mass gluinos ranging from 200 to 600\GeV,
	 and one for high-mass gluinos covering the remainder of the mass range up
	 to 1500\GeV. The solid red lines show the NLO+NLL
	 predictions~\cite{susyxsec1, susyxsec2, susyxsec3, susyxsec4, susyxsec5},
	 and the dashed red lines give
	 the corresponding one-standard-deviation uncertainty bands~\cite{susyxsecuncert}.}
\label{fig:Limits}
\end{figure}

Upper limits are placed on the cross section times branching fraction for the
production of three-jet resonances. A modified-frequentist approach, using
the $\mathrm{CL}_\mathrm{s}$~\cite{Junk:1999kv,Read:2002hq}
technique and a profile likelihood as the test statistic, is employed.
Limits are calculated with the frequentist
asymptotic calculator implemented in the \textsc{RooStats}~\cite{cowan, RooStats} package.
The full $\mathrm{CL}_\mathrm{s}$ calculations give similar limits within a few percent, and closure tests 
where a fixed signal is injected, yield consistent coverage.
The observed and expected 95\%
confidence level (CL) upper limits on the gluino pair-production cross section
times branching fraction as a function of gluino mass are presented in
Fig.~\ref{fig:Limits}. The solid red lines in the figure show the next-to-leading-order (NLO) plus
next-to-leading-logarithm (NLL)
cross sections for gluino pair production~\cite{susyxsec1, susyxsec2, susyxsec3, susyxsec4, susyxsec5},
and the dashed red lines indicate the corresponding
one-standard-deviation ($\sigma$) uncertainties,
which range between 15\% and 43\%.
To quote final results, we use the points where
the $-1\sigma$-uncertainty curve for the NLO+NLL cross section crosses the
expected- and observed-limit curves.
We additionally quote the result where the central theoretical curve
intersects the limit curves.

The production of gluinos undergoing RPV decays into light-flavour jets is
excluded at 95\% CL for gluino masses below 650\GeV, with a less conservative exclusion of 670\GeV based 
upon the theory value at the central scale. The respective expected limits are
755 and 795\GeV.
These results
extend the limit of 460\GeV~\cite{gluino2011}
obtained with the 7\TeV CMS dataset. Gluinos whose decay includes a heavy-flavour jet are
excluded for masses between 200 and 835\GeV, which is the most stringent mass limit
to date for this model of RPV gluino decay, with the less conservative exclusion up to 855\GeV
from the central theoretical value. The respective expected limits are 825 and 860\GeV.
While a smaller phase space is probed in the heavy-flavour search, the limits
extend to higher masses because of the reduction of the background.
\section{Summary}
A search for hadronic resonance production in \Pp\Pp\ collisions at a centre-of-mass
energy of 8\TeV has been conducted by the CMS experiment at the LHC with a data
sample corresponding to an integrated luminosity of 19.4\fbinv. The approach is
model independent, with event selection criteria optimised using the RPV
supersymmetric model
for gluino pair production in a six-jet final state. Two different scenarios for this
RPV decay have been considered: gluinos decaying exclusively to light-flavour jets,
and gluinos decaying to one b-quark jet and two light-flavour jets, with the assumption in
both cases of a 100\% branching fraction for gluinos decaying to quark jets. 
Methods based on data have been used to derive estimates of background from SM multijet processes.
Events with
high jet multiplicity and a large scalar sum of jet \pt have been analysed for the
presence of signal events, and no deviation has been found between the standard
model background expectations and the measured mass distributions. The production of gluinos undergoing RPV decay into
light-flavour jets has been excluded  at the 95\% CL for masses below 650\GeV.
Gluinos that include a heavy-flavour jet in their decay have been excluded at 95\% CL for masses between 200
and 835\GeV, which is the most stringent limit to date for this model of gluino
decay.

\section*{Acknowledgements}
We congratulate our colleagues in the CERN accelerator departments for the excellent
performance of the LHC and thank the technical and administrative staffs at CERN and at
other CMS institutes for their contributions to the success of the CMS effort. In addition,
we gratefully acknowledge the computing centres and personnel of the Worldwide LHC
Computing Grid for delivering so effectively the computing infrastructure essential to
our analyses. Finally, we acknowledge the enduring support for the construction and operation of the LHC and the CMS detector provided by the following funding agencies: BMWF and FWF (Austria); FNRS and FWO (Belgium); CNPq, CAPES, FAPERJ, and FAPESP (Brazil); MES (Bulgaria); CERN; CAS, MoST, and NSFC (China); COLCIENCIAS (Colombia); MSES (Croatia); RPF (Cyprus); MoER, SF0690030s09 and ERDF (Estonia); Academy of Finland, MEC, and HIP (Finland); CEA and CNRS/IN2P3 (France); BMBF, DFG, and HGF (Germany); GSRT (Greece); OTKA and NKTH (Hungary); DAE and DST (India); IPM (Iran); SFI (Ireland); INFN (Italy); NRF and WCU (Republic of Korea); LAS (Lithuania); CINVESTAV, CONACYT, SEP, and UASLP-FAI (Mexico); MBIE (New Zealand); PAEC (Pakistan); MSHE and NSC (Poland); FCT (Portugal); JINR (Dubna); MON, RosAtom, RAS and RFBR (Russia); MESTD (Serbia); SEIDI and CPAN (Spain); Swiss Funding Agencies (Switzerland); NSC (Taipei); ThEPCenter, IPST, STAR and NSTDA (Thailand); TUBITAK and TAEK (Turkey); NASU (Ukraine); STFC (United Kingdom); DOE and NSF (USA).

\bibliography{auto_generated}   

\cleardoublepage \appendix\section{The CMS Collaboration \label{app:collab}}\begin{sloppypar}\hyphenpenalty=5000\widowpenalty=500\clubpenalty=5000\textbf{Yerevan Physics Institute,  Yerevan,  Armenia}\\*[0pt]
S.~Chatrchyan, V.~Khachatryan, A.M.~Sirunyan, A.~Tumasyan
\vskip\cmsinstskip
\textbf{Institut f\"{u}r Hochenergiephysik der OeAW,  Wien,  Austria}\\*[0pt]
W.~Adam, T.~Bergauer, M.~Dragicevic, J.~Er\"{o}, C.~Fabjan\cmsAuthorMark{1}, M.~Friedl, R.~Fr\"{u}hwirth\cmsAuthorMark{1}, V.M.~Ghete, N.~H\"{o}rmann, J.~Hrubec, M.~Jeitler\cmsAuthorMark{1}, W.~Kiesenhofer, V.~Kn\"{u}nz, M.~Krammer\cmsAuthorMark{1}, I.~Kr\"{a}tschmer, D.~Liko, I.~Mikulec, D.~Rabady\cmsAuthorMark{2}, B.~Rahbaran, H.~Rohringer, R.~Sch\"{o}fbeck, J.~Strauss, A.~Taurok, W.~Treberer-Treberspurg, W.~Waltenberger, C.-E.~Wulz\cmsAuthorMark{1}
% ...
\vskip\cmsinstskip
% ...
1:~~Also at Vienna University of Technology, Vienna, Austria\\
2:~~Also at CERN, European Organization for Nuclear Research, Geneva, Switzerland\\
% ...

\end{sloppypar}
\end{document}